\documentclass[twocolumn,showpacs,preprintnumbers,amsmath,amssymb,nofootinbib]{revtex4}
\usepackage{graphicx,color}
\usepackage{amsmath,amssymb}
\usepackage{bm}

\newcommand\nn{\nonumber \\}
\newcommand{\psibar}{\mbox{${\bar \psi}$}}

\newcommand{\vcp}{\mbox{$\bm{p}$}}

\newcommand{\vbk}{\mbox{$\bm{k}$}}

\newcommand{\valph}{\mbox{$\bm{\alpha}$}}

\def\sla{\slash{\!\!\!}}

\newcommand{\lk}{\left(}
\newcommand{\rk}{\right)}

\newcommand{\ldk}{\left[}

\newcommand{\rdk}{\right]}
\newcommand\beq{ \begin{eqnarray} }
\newcommand\eeq{ \end{eqnarray} }

\begin{document}

%\preprint{}
\title{Spin polarized phases in strongly interacting matter: 
interplay between axial-vector and tensor mean fields}
\author{Tomoyuki Maruyama}
\email{maruyama.tomoyuki@nihon-u.ac.jp} 
\affiliation{College of Bioresource Sciences, Nihon University, Fujisawa 252-8510, Japan}
\author{Eiji Nakano}
\email{e.nakano@kochi-u.ac.jp}
\affiliation{Department of Physics, Kochi University, Kochi 780-8520, Japan}
\author{Kota Yanase}
\email{yanase@nuclei.th.phy.saitama-u.ac.jp} 
\affiliation{Department of Physics, Saitama University, Saitama City 338-8570, Japan}
\author{Naotaka Yoshinaga}
\email{yoshinaga@phy.saitama-u.ac.jp}
\affiliation{Department of Physics, Saitama University, Saitama City 338-8570, Japan}

\date{\today}

\begin{abstract}
The spontaneous spin polarization  of  strongly interacting matter
 due to axial-vector and tensor type interactions is studied at zero
 temperature and high baryon-number densities.  
We start with the mean-field Lagrangian for the axial-vector and tensor interaction channels, 
and find in the chiral limit that the spin polarization due to the tensor mean field ($U$) takes place first as the density increases for sufficiently strong coupling constants, 
and then that due to the axial-vector mean field  ($A$) emerges 
in the region of finite tensor mean field. 
This can be understood that making the axial-vector mean field finite requires 
a broken chiral symmetry somehow, 
which is achieved by the finite tensor mean field in the present case.  
It is also found from symmetry argument that 
there appear the type I (II) Nambu-Goldstone modes with a linear (quadratic) dispersion 
in the spin polarized phase with $U\neq0$ and $A=0$  ($U\neq0$ and $A\neq0$), 
although these two phases exhibit the same symmetry breaking pattern. 
\end{abstract} 

%\pacs{}
%\keywords{}

\maketitle

%\tableofcontents

\section{Introduction}

Discovery of the magnetars,
the neutron stars with strong magnetic field of ${\mathcal O}\lk 10^{15}\rk$~G, revives 
the important question about the origin of the strong magnetic field \cite{mag3,mag,ThDr95,MagRev}. 
Recent studies have also revealed 
that magnetars possess 
not only the poloidal magnetic field but also 
the toroidal one 
whose strength is about $10^2$ times larger than the former \cite{Braithwaite1,makishima14}.
There are several arguments about how such strong magnetic fields are generated
and survive in the evolution of neutron stars \cite{MagRev}, 
but definite conclusion is yet to be made. 

The spontaneous spin polarization or magnetization of the strongly
interacting matter is one of important issues
in the relation to such strong magnetic field. 
As an earlier work, Tatsumi \cite{tat00} suggested a possibility of a ferromagnetic 
transition in quark matter interacting via one-gluon-exchange (OGE) force 
and showed that the maximum magnetic field can reach
$B \sim O(10^{15-17}{\rm G})$ when the magnetar is a quark star.

In general a ferromagnetic nature of dense matter 
manifests itself when a spin polarization of charged fermions, i.e., baryons or quarks, occurs collectively by their interactions, 
so that the {\it spin} degrees of freedom is a key ingredient.  
In the relativistic framework we can consider 
two types of spinor bilinear form as the {\it spin} density operator \cite{Maruyama:2000cw}: 
One is a spatial component of  the axial-vector (AV) current operator, 
$\psi^\dagger \Sigma_i \psi  (\equiv -\psibar \gamma_5 \gamma_{i}\psi )$, 
and the other  is that of the tensor (T) operator, 
$\psi^\dagger \gamma^0 \Sigma_i \psi (\equiv -\frac{\epsilon_{ijk}}{2} \psibar \sigma_{jk} \psi)$,
with $\psi$ being the Dirac field. 
These two become equivalent to each other in the non-relativistic limit, 
while they are quite different in the ultra-relativistic limit (massless limit) \cite{Maruyama:2000cw}. 
When the expectation value of AV and/or T operators becomes finite, 
the spin polarization (SP) is realized.   
In fact the T expectation value, 
$\langle \psi^\dagger \gamma^0 \Sigma_i \psi\rangle$, directly reflects
the magnetic effect through the electro-magnetic coupling 
$\frac{Q}{2m}\psibar \sigma_{\mu \nu} \psi F^{\mu \nu}$ (Gordon decomposition) for a particle with charge $Q$ and mass $m$, 
which is reminiscent of the Ising model under an external magnetic field. 
On the other hand, 
the finite AV expectation value,  $\langle \psi^\dagger \Sigma_i \psi\rangle$, 
leads to a spin polarization, 
since the spatial components of the AV current 
correspond to the generators of rotations  
in the spinor representation of the Lorentz group. 

In the previous studies we have entirely relied on the mean-field approach, 
where we consider the AV and T channel interactions and make mean fields for them. 
As origin of such interactions, 
the AV-type interactions in quark matter can be derived from,  
e.g.,  the perturbative OGE interaction 
using the Fierz transformation, 
while the T-type interactions are expected to appear via the non-perturbative effects of QCD 
as seen in hadron-hadron effective interactions \cite{Ring1}. 
Then, the  effective models of QCD should include both types of interactions. 
So far the interplay between the spin polarization and other phases expected to appear in high baryon-number density region have been studied, 
which includes  the co-existence of the SP and the color super conductivity 
\cite{nak03A,Tsue:2012nx,Tsue:2014tra,Matsuoka:2016ufr}, 
the spatially homogeneous 
\cite{Maedan:2006ib, TPPY12, Matsuoka:2016spx, Maruyama:2017mqv, Morimoto:2018pzk} 
and inhomogeneous chiral condensations \cite{NT05, Yoshiike:2015tha}. 
From these studies it is found that 
the AV- and T-type mean fields are affected differently by the dynamical chiral symmetry breaking: 
When the dynamical quark mass is zero, i.e., the chiral symmetry is restored, 
the AV-type spin polarized phase cannot appear, but the T-type one can do. 
For instance, in the NJL type effective models, it has been demonstrated that 
the AV-type spin polarized phase can appear only in a narrow density region 
just inside the chiral condensed phase 
\cite{Maedan:2006ib,Maruyama:2017mqv,Morimoto:2018pzk},
while the T-type spin polarized phase can exist in even higher density regions  
regardless whether the dynamical quark mass is finite or not 
\cite{Maruyama:2000cw,TPPY12,Matsuoka:2016spx,Maruyama:2017mqv}. 
As will be shown below, 
this is because the T-type condensation itself breaks the chiral symmetry 
while the AV-type one respects them.
So far we have not known  the spin polarized phase of systems  
including both the AV- and T- type interactions simultaneously, 
which is expected to exhibit new features of the SP. 
Thus, in the present study, we investigate the interplay between them on the same footing,  and figure out the phase structure in terms of 
the coupling strengths of AV and T channels, and 
of the baryon-number chemical potential at zero temperature in the chiral limit. 

The paper is organized as follows: 
in the next section we formulate a mean-field Lagrangian 
with AV and T mean fields and its thermodynamic potential 
at finite baryon-number density and at zero temperature. 
In the section III, 
after discussion of the relation between the spin polarization and the chiral symmetry,  
we search out the potential minimum 
to find the phase structure in the space of coupling constants. 
In the section IV, we demonstrate the change of phase structure with the chemical potential, employing a chiral model, and show that there appear two kinds of Nambu-Goldstone modes 
depending on the finiteness of the AV mean field.  
The last section is devoted to summary and outlook. 
%%%%%%%%%%%%%%%%%%%%%%%%%%%%%%%%%%%%%
\section{Mean-field approximation and thermodynamic potential at $T=0$ and $\mu\neq 0$}
%%%%%%%%%%%%%%%%%%%%%%%%%%%%%%%
In this section we briefly explain our formalism, 
which holds the flavor $SU(2)$ and the color $SU(3)$ symmetry.
In addition, we consider only the spin-isospin saturated quark matter.  
We start with a  general Lagrangian density 
including the spatial parts of AV and T fields, 
\begin{equation}
L=\bar{\psi}\lk i \sla{\partial} - m \rk \psi
 + A_i \bar{\psi} \gamma_5 \gamma_i \psi
 + U_{ij} \bar{\psi} \sigma_{ij}  \psi -\frac{A_i^2}{g_A} -\frac{U_{ij}^2}{g_U}, 
\label{mfL}
\end{equation}
where $\psi$ is the quark field,  $m$ the quark mass, 
 $g_{A,U}$ effective coupling constants of AV and T channels, 
 $A_i=g_A \langle  \bar{\psi} \gamma_5 \gamma_i \psi\rangle$, and 
 $U_{ij}=g_U \langle  \bar{\psi} \sigma_{ij} \psi\rangle$. 
This Lagrangian is applicable for effective strong interactions included in 
e.g., NJL model, linear sigma model, and quark-quark interactions with screened gluons. 
In the isospin saturated system the isospin dependent terms do not contribute to the mean field, 
and we omit them in the above Lagrangian. 

Here, we assume only the third components of the mean fields, 
$A_3(=A)$ and $U_{12}(=U)$, to be nonzero, and obtain 
the Dirac equation for the spinor $u(\vbk,s)$ with momentum $\vbk=(k_x,k_y,k_z)$ and spin $s$, 
\begin{equation}
\left[ \valph \cdot \vbk + m + \Sigma_z A + \beta \Sigma_z U
\right] u(\vbk,s) = \varepsilon_{k,s} u(\vbk,s) .
\end{equation}
The single particle energy $\varepsilon_{k,s}$ 
is obtained as a solution of the  characteristic equation for $\varepsilon$, 
\begin{eqnarray}
&& 
(\varepsilon^2 - E_k^2)^2 - 2 \varepsilon^2 (A^2 + U^2) 
- 8  m A U \varepsilon - 2 m^2 (A^2 + U^2) 
\nonumber \\ && \qquad
+ (A^2 - U^2)^2 -2 (A^2 - U^2) (k_z^2 - k_t^2)  =  0 ,
\label{detAT} 
\end{eqnarray}
where  $E_k=\sqrt{\vbk^2+m^2}$, and $k_t=\sqrt{k_x^2+k_y^2}$ 
is the magnitude of the transverse momentum normal to the polarization direction $z$. 
Since the above equations include the quark mass, 
it is easy to extend the present formulation to involve the chiral condensation 
in the same mean-field approximation.  
Nevertheless, 
we are interested in the high density region where the chiral condensation has already gone, 
and intend to make discussion about the chiral symmetry transparent, 
so we take the chiral limit $ m \rightarrow 0$ in the following calculations.  

The thermodynamics potential is given by 
\begin{equation}
\Omega\ldk A, U, \mu \rdk 
=
\Omega_F\ldk A, U, \mu \rdk 
+
\Omega_D\ldk A, U, \mu \rdk 
+ \frac{A^2}{g_A}+\frac{U^2}{g_U}, 
\label{thermPot1}
\end{equation}
where $\Omega_F$ is the contribution from the matter component 
up to the Fermi surfaces, 
and $\Omega_D$ from the Dirac sea, given respectively by 
\beq
\Omega_F\ldk A, U, \mu \rdk 
&=&
N_d \sum_{s=\pm 1} \int_k 
\lk\varepsilon_{k,s}-\mu\rk 
\theta\lk \mu -  \varepsilon_{k,s} \rk, 
\\
\Omega_D\ldk A, U, \mu \rdk 
&=&
- N_d \sum_{s=\pm 1} \int_k \varepsilon_{k,s},
\eeq
where the abbreviated notation $\int_k \equiv \int {\rm d}k^3$ is used, 
$N_d$ is the degeneracy factor, 
$\mu$ the chemical potential, and 
the single particle energies $\varepsilon_{k,s}$ in the chiral limit, 
\begin{equation}
\varepsilon_{p,s}
=
\sqrt{k_z^2+k_t^2+A^2+U^2+2s\sqrt{k_z^2A^2+k_t^2U^2+A^2U^2}}. 
\label{spe}
\end{equation}
Note that the splitting of the energy spectrum by $s=\pm 1$ corresponds to the spin polarization 
due to the mean fields, and the magnitude of the mean-fields is determined mainly 
by the matter contribution $\Omega_F$ as many-body effects, 
unlike the chiral condensation that comes from the Dirac sea contribution. 
 
Although the above formulation can be applied to the both of hadronic and quark matters, 
we use $N_d=6$ (two flavors times three colors), 
and $\mu$ to be the quark chemical potential from now on. 
Also, from results of the preceding studies that 
the spin polarization due to the AV or T mean fields  
is favored at higher densities where the chiral condensation already diminishes, 
we will neglect the Dirac sea contribution $\Omega_D$ 
of the thermodynamic potential in what follows. 

\subsection{Fermi surfaces}
\begin{figure}[htbp]
  \begin{center}
\includegraphics[height=8.0cm]{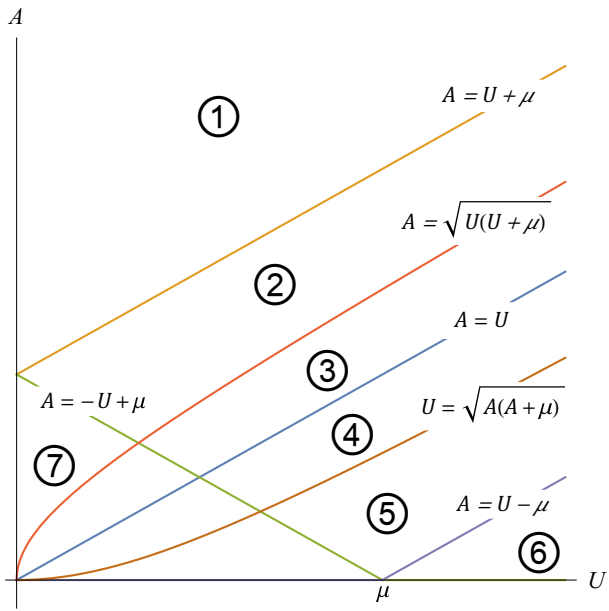}
%    \begin{tabular}{cc}
% \resizebox{50mm}{!}{\includegraphics{nemo1.eps}} & 
% \resizebox{50mm}{!}{\includegraphics{nemo2.eps}} \\ 
% \resizebox{50mm}{!}{\includegraphics{nemo2.eps}} & 
% \resizebox{50mm}{!}{\includegraphics{nemo1.eps}} \\ 
%    \end{tabular}
    \caption{ \textcircled{1}- \textcircled{6}  correspond to different topologies of Fermi surfaces for $s=-1$ as shown in Fig.~\ref{fig2}, 
     and the Fermi surface for $s=1$ becomes finite only in the region of \textcircled{7}: $A<-U+\mu$.   }
    \label{fig1}
  \end{center}
\end{figure}

\begin{figure*}[htbp]
  \begin{center}
    \begin{tabular}{ccc}
 \resizebox{50mm}{!}{\includegraphics{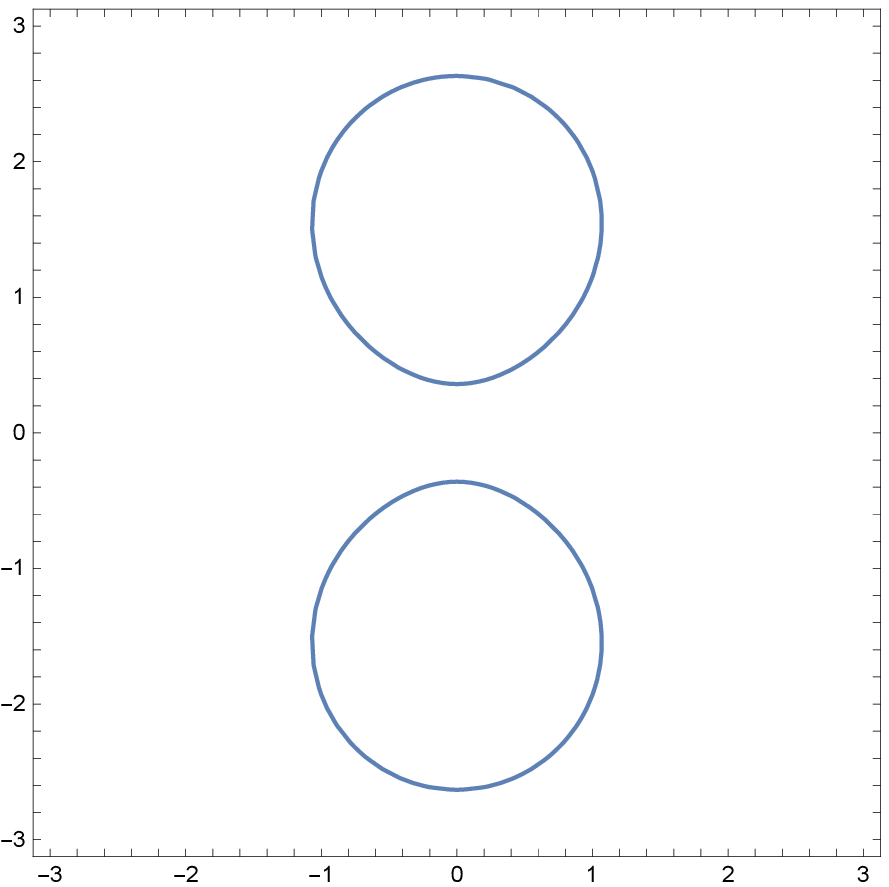}} & 
 \resizebox{50mm}{!}{\includegraphics{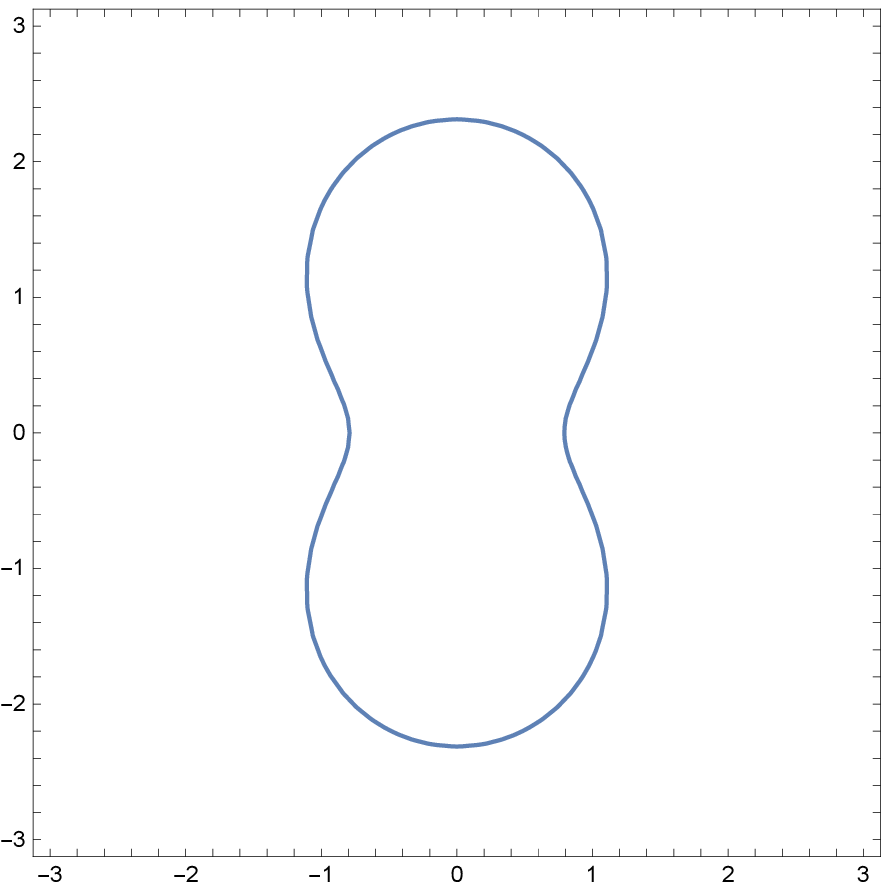}} & 
 \resizebox{50mm}{!}{\includegraphics{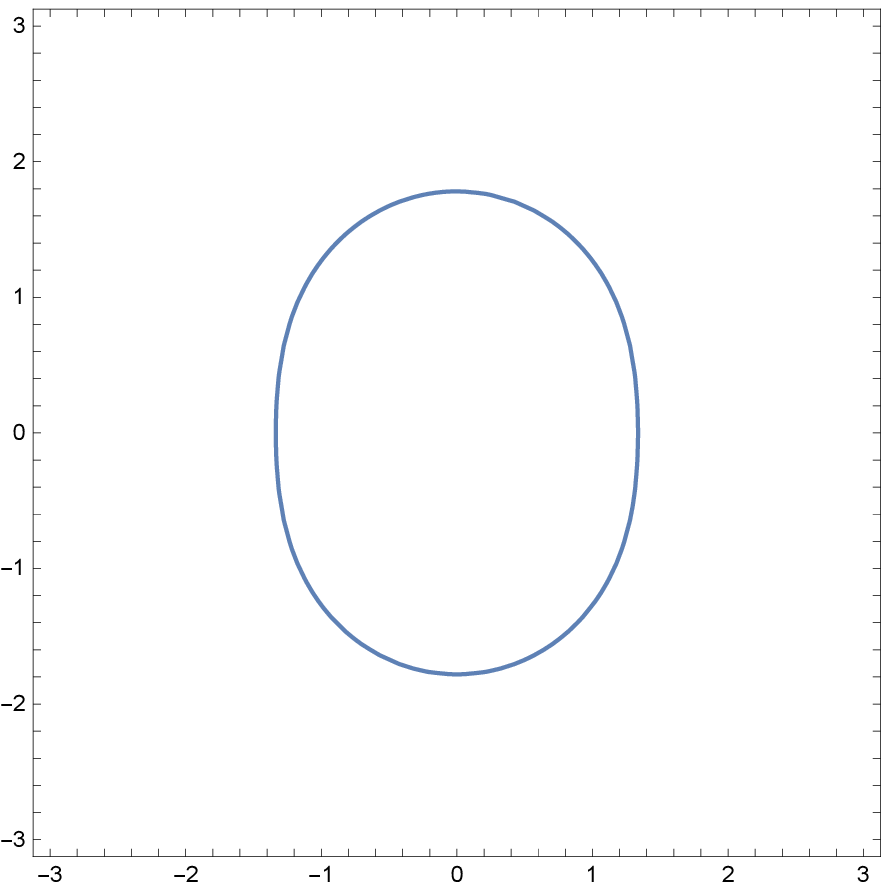}} \\ 
 \resizebox{50mm}{!}{\includegraphics{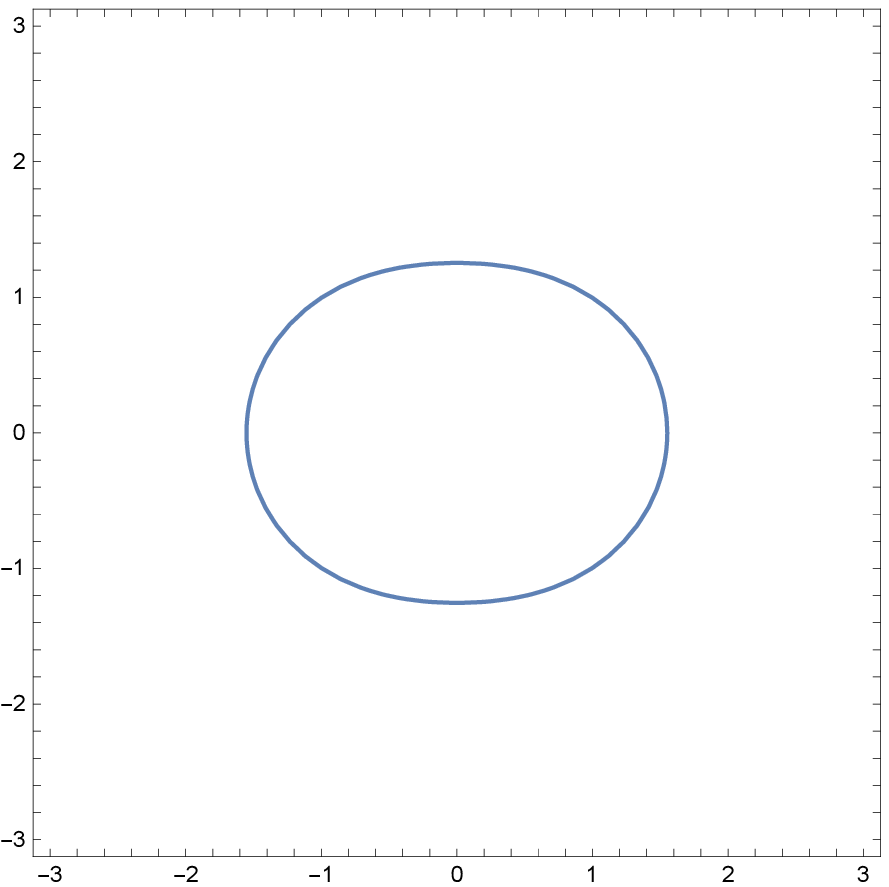}} & 
 \resizebox{50mm}{!}{\includegraphics{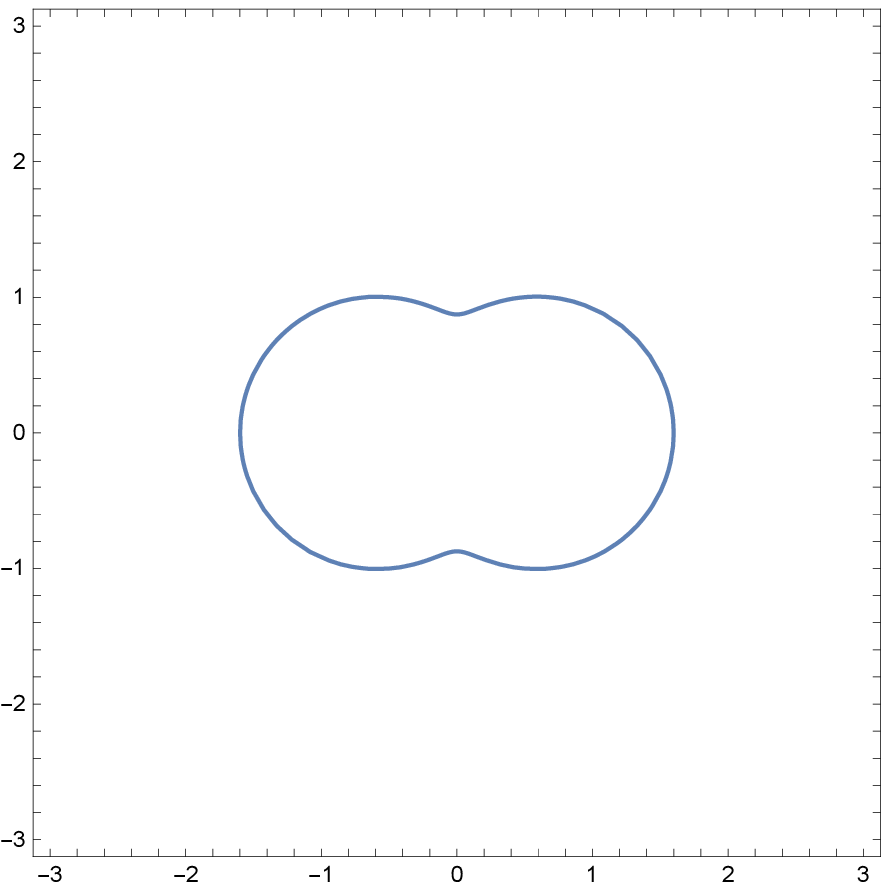}} & 
 \resizebox{50mm}{!}{\includegraphics{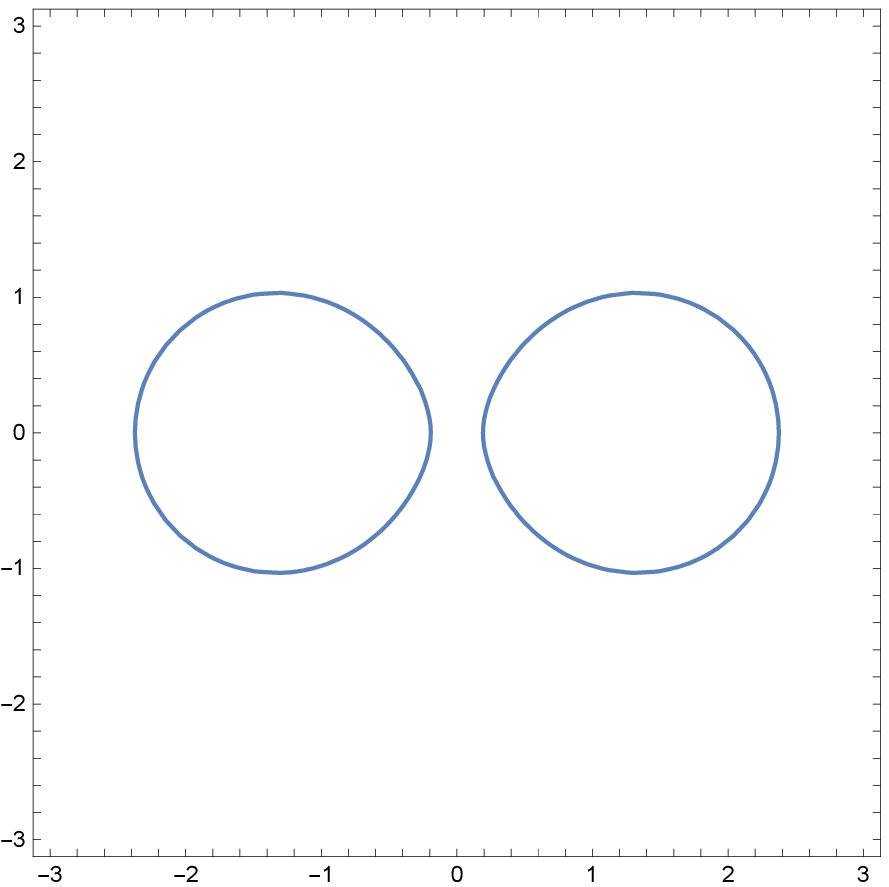}} \\ 
    \end{tabular}
    \caption{ Cross sections of Fermi surfaces in $k_y-k_z$ plane for $s=-1$. 
    Ordinate is $k_z$ and abscissa is $k_y$. 
    In the top row from left to right \textcircled{1} $\rightarrow$ \textcircled{3}, 
    and in the bottom row from left to right \textcircled{4} $\rightarrow$ \textcircled{6}. 
    Note that the Fermi surface is symmetric under the rotation around $k_z$ axis, so the last one is a doughnut in full three dimension.}
    \label{fig2}
  \end{center}
\end{figure*}

Since the single particle energies Eq.~(\ref{spe})  are split and deformed by the mean-fields, 
we take care of the topology of Fermi surfaces in the calculation of $\Omega_F$, 
especially for the $s=-1$ branch.  
The modification of the Fermi surface for $s=-1$ is classified for values of $A, U$, and $\mu$ 
in Fig.~\ref{fig1}, and corresponding Fermi surfaces are shown in Fig.~\ref{fig2}. 
%%%%%%%%%%%%%%%%%%%%%%%%%%%%%%%%%%%%%%%%%%%%%%%%%%%%%%%%%%%%%
%%%%%%%%%%%%%%%%%%%%%%%%%%%%%%%%%%%%%%%%%%%%%%%%%%%%%%%%%%%%%%
%%%%%%%%%%%%%%%%%%%%%%%%%%%%%%%%%%%%%%%%%%%%%%%%%%%%%%%%%%%%%
%%%%%%%%%%%%%%%%%%%%%%%%%%%%%%%%%%%%%%%%%%%%%%%%%%%%%%%%%%%%%%
\subsection{Thermodynamic potential $\Omega_F$ from Fermi seas }
The integral of $\Omega_F$ in the transverse direction ($y=k_x^2+k_y^2$) 
can be done analytically: 
\beq
&&\int_k \varepsilon_{k,s} \theta(\mu-\varepsilon_{k,s})
=
\frac{1}{4\pi^2} 
\int  {\rm d}k_z \int_{y_1}^{y_2}\frac{ {\rm d}y}{2} 
\varepsilon_{k,s}
%%%%%%%
\nn 
&=&
\frac{1}{8\pi^2} 
\int_{z_1}^{z_2}  {\rm d}k_z
\lk \frac{\varepsilon_{k,s}^2}{6}
+\frac{y+k_z^2+A^2-U^2}{2}    \rk \varepsilon_{k,s}
\nn
&&
+s\frac{k_z^2\lk A^2-U^2\rk }{U}
\nn
&&
\times \left. \ln\lk \sqrt{yU^2+A^2(k_z^2+U^2)}+sU^2+U\varepsilon_{k,s} \rk
\right|_{y_1}^{y_2} 
\label{omegaF}
\eeq
The integral ranges $z_{1,2}$ and $y_{1,2}$ are determined 
in accordance with the Fermi surfaces for $s=\pm 1$, 
and details of the calculations are given in Appendix~A. 

%%%%%%%%%%%%%%%%%%%%
\section{Phase structure and discussion}
%%%%%%%%%%%%%%%%%%%%
The present study aims to figure out the phase structure in the space of coupling strengths and the chemical potential by searching out the minimum points 
of the thermodynamic potential $\Omega(A,U)$.  
To this end it is heuristic to start with argument on properties of the mean fields: 
The spin polarization by $A\neq0$ never occurs at $U=0$ in the chiral limit, 
that is, 
the $\Omega[A,U]$ is always stable against $A$ fluctuations at the origin in the $A$-$U$ space. 
This is because in general AV-type  mean field, 
appearing in the form of a mean-field interaction term 
$A_{ia} \bar{\psi}\gamma_5\gamma_i \tau_a \psi$ 
with $\tau_{a=0,1,2,3}$ being the identity or Pauli matrices in the isospin space, 
can be  eliminated by a local chiral transformation 
$\psi \rightarrow e^{i\gamma_5A_{ia} \tau_a x_i}\psi$  
through the derivative term \cite{Namb04}, 
which costs zero energy in the chiral limit (by the redefinition of the spinor field). 
In other words, only in the case that the chiral symmetry is broken, 
the net expectation value of the AV mean field can be generated 
%%%%
\footnote{
Impacts of the axial anomaly to the axial-vector mean field at finite densities is neglected in this study.}.
%%%%
In the present case, 
a finite $U$ breaks the time reversal symmetry like in an external magnetic field, 
and the $L$-$R$ symmetry as well, thus can invoke a finite $A$ for sufficiently strong couplings. 
From these observations we set the strategy to get the phase structure as follows: 
We first find out the minimum point, $U_{\rm min}$, of  the potential on the $U$ axis ($A=0$), 
and then check if the 2nd order derivative (curvature) in the $A$ direction 
is negative $\partial^2 \Omega /\partial A^2 <0$ or 
positive $\partial^2 \Omega /\partial A^2 >0$
at the minimum point.   
In the former case  the phase with $(A\neq 0, U\neq0)$ can be realized, 
while in the latter case a possible phase corresponds to $(A=0, U\neq0)$, 
which, however, can be a local minimum.  
We have checked such situations,  
and found no global minimum away from the $U$ axis  in the present approximation. 

As shown below, 
there exist some key points of the coupling constants, 
which determine the topology of the Fermi surfaces 
and signs of the potential curvatures on the $U$ axis.
We give these relations 
at $N_d=6$ in the following:
\begin{enumerate}
\item 
For $g_U>g_{U {\rm crit}}\equiv
\frac{4\pi^2}{N_d \mu^2}=\frac{6.57974}{\mu^2}$, 
the potential curvature to the $U$ direction becomes negative,
$\partial^2 \Omega[A,U]\partial U^2 < 0$,  
at $U=A=0$. 
\item 
For $g_U\ge g_{U1} \equiv\frac{24\pi}{N_d \mu^2}=\frac{12.5664}{\mu^2}$, 
$U_{\rm min}\ge \mu $ 
where  $U_{\rm min}$ is the minimum point of the potential on the $U$ axis. 
\item 
For $g_U\ge g_{U2} \equiv \frac{12.069141}{\mu^2}$, 
$U_{\rm min}\ge U_{\rm cri}=0.959993\mu$, 
which is determined by $\partial_U \partial_A^2 \Omega[0,U_{\rm cri}]=0$. 
\item 
$g_{A1}=\frac{12.4077}{\mu^2}$. 
Only if $g_A\ge g_{A1} \equiv \frac{74.4462}{N_d \mu^2}=\frac{12.4077}{\mu^2}$, 
the potential curvature can be negative, 
$\partial^2 \Omega/ \partial A^2 < 0$ at $A=0$.
\item 
When $g_{A2}=\frac{12.5664}{\mu^2}$, 
$\partial^2 \Omega / \partial A^2 =0$ at $U=\mu$ and $A=0$. 
\end{enumerate}
From these points we obtained the boundary of the spin polarized phases 
as shown in Fig.~\ref{fig3}.  
Its detailed derivation is given in Appendix~C. 
%%%%%%%%%%%%%%%%%%%%%%%%%%%%%%%%%%%%%%%%%%%%%%%%%%%%%%%%%%%%%
\begin{figure}[htbp]
	\begin{center}
		\includegraphics[height=8.0cm]{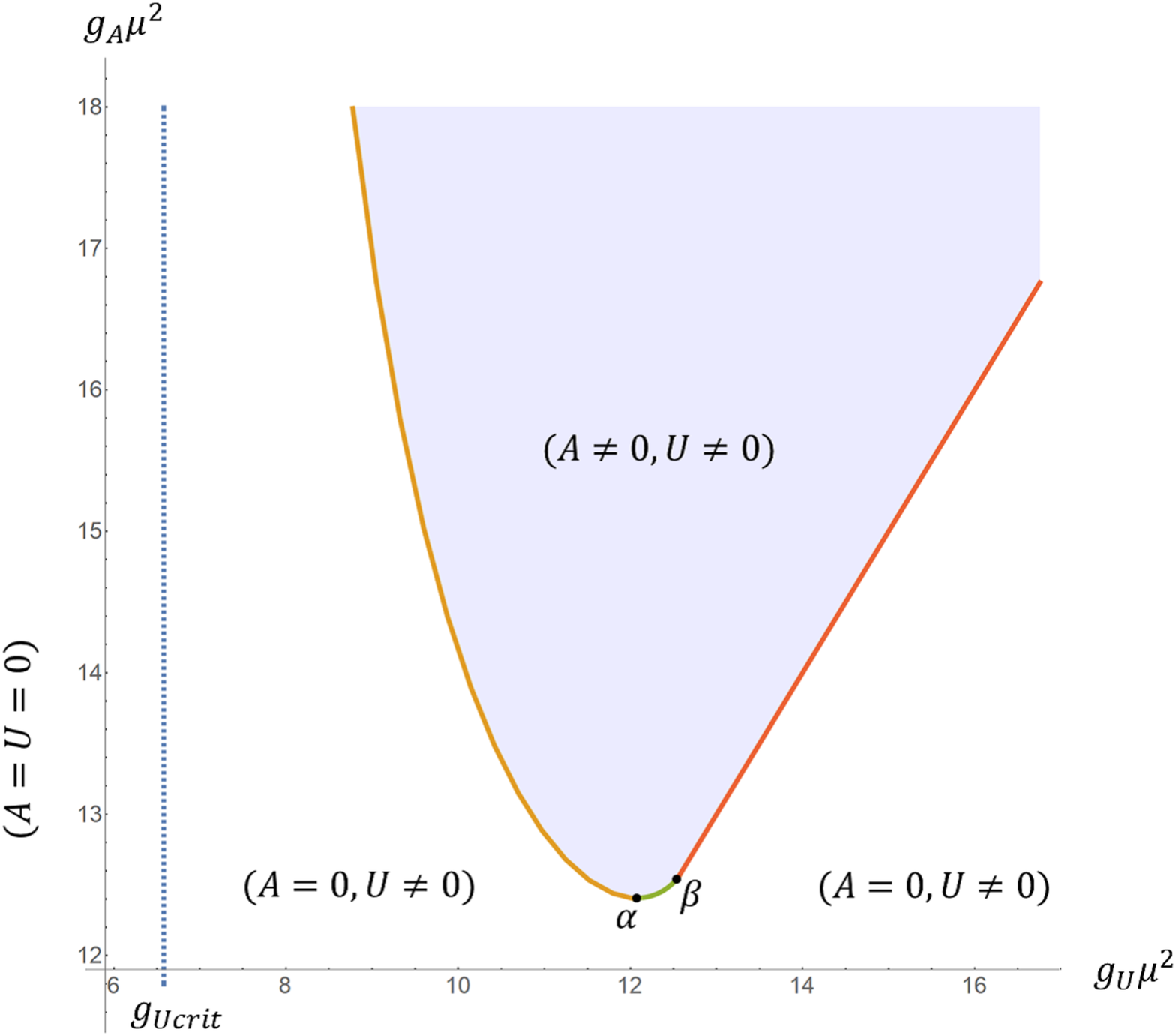}
		%    \begin{tabular}{cc}
		% \resizebox{50mm}{!}{\includegraphics{nemo1.eps}} & 
		% \resizebox{50mm}{!}{\includegraphics{nemo2.eps}} \\ 
		% \resizebox{50mm}{!}{\includegraphics{nemo2.eps}} & 
		% \resizebox{50mm}{!}{\includegraphics{nemo1.eps}} \\ 
		%    \end{tabular}
		\caption{ Phase structure in the plane of axial-vector and tensor couplings, 
$g_A$ and $g_U$, normalized by the chemical potential $\mu$. 
The shaded region bounded from below by the solid curve corresponds 
to the $(A\neq 0, U\neq 0)$ phase. 
The other region is separated by the dotted line ($g_U=g_{Ucrit}$): 
$g_U > g_{U {\rm Crit}}$ corresponds to the $(A= 0, U\neq 0)$ phase, 
and $g_U \le g_{U {\rm Crit}}$ to the normal one $(A= U= 0)$.
Characteristic points are indicated by $\alpha=(g_{U2}, g_{A1})$ and $\beta=(g_{U1}, g_{A2})$.}
		\label{fig3}
	\end{center}
\end{figure}
%%%%%%%%%%%%%%%%%%%%%%%%%%%%%%%%%%%%%%%%%%%%%%%%%%%%%%%%%%%%%%
The phase structure is summarized as follows: 
The normal phase $(A=U=0)$ appears for small tensor couplings 
$g_U \le g_{U {\rm Crit}}$. 
For $g_U > g_{U {\rm Crit}}$ the T mean field becomes always finite, 
but the finiteness of the AV mean field depends on the strength of the coupling $g_A$,   
which is bounded from below, i.e., $g_A > g_{A1}=12.4077/\mu^2$. 
In strong coupling regions where $g_A, g_U \ge 12.5664/\mu^2$,  
the two spin polarized phases  are separated by the straight line $g_A=g_U$.

\section{application to an extended NJL model}
In the microscopic description of the strong interaction, 
perturbative vector-vector type interactions, such as the single-gluon exchange interaction,  
do not generate T-type interactions even after the Fierz transformation, 
while effective models of the strong interaction are able to accommodate them as non-perturbative effects, 
e.g., the instanton induced interaction 
\cite{Schafer:1996wv,Carter:1998ji}. 
As an application we employ an extended NJL model as an effective chiral model 
\cite{Hatsuda:1994pi,Maruyama:2000cw}, 
\beq
L&=&\bar{\psi} \sla{\partial}\psi
-G_s \ldk \lk\bar{\psi} \psi \rk^2
+\lk\bar{\psi} i\gamma_5\tau_a\psi \rk^2\rdk 
\nn 
&&
-G_A\ldk \lk\bar{\psi} \gamma_\mu \psi \rk^2
+\lk\bar{\psi} \gamma_5\gamma_\mu \tau_a\psi \rk^2\rdk 
\nn 
&&
-G_U\ldk \lk\bar{\psi} \sigma_{\mu\nu}\psi \rk^2
+\lk\bar{\psi} i\gamma_5\sigma_{\mu\nu} \tau_a  \psi \rk^2\rdk 
\label{enjl}
\eeq
Using the Fierz transformation \cite{Klevansky1} 
we can single out the relevant interaction terms (exchange channels) 
which are to be mean fields, 
showing only the spatial components of AV and T terms explicitly, 
\begin{equation}
L = \bar{\psi} \sla{\partial}\psi +\frac{g_A}{2}\lk\bar{\psi} \gamma_5 \gamma_i \psi \rk^2
+\frac{g_U}{2}\lk\bar{\psi} \sigma_{ij} \psi \rk^2+\cdots, 
\end{equation}
where 
$\frac{g_A}{2}=%-\frac{1}{8}G_s -\frac{3}{8}G_s -\frac{1}{4}G_A -\frac{3}{4}G_A= 
-\frac{1}{2}G_s -G_A$, and  
$\frac{g_U}{2}=%\frac{1}{16}G_s -\frac{3}{16}G_s  -\frac{3}{4}G_U -\frac{1}{4}G_U=
-\frac{1}{8}G_s -G_U$. 
%%%%%%%%%%%%%%%%%%%%%%
Taking the mean-field approximation as 
$A=A_3=g_A \left\langle \bar{\psi} \gamma_5 \gamma_3 \psi \right\rangle$, 
and $U=U_{12}=g_U\left\langle\bar{\psi} \sigma_{12} \psi \right\rangle$ 
for the spin polarization to $z$ direction,  
consistently with vanishing others $A_1=A_2=U_{13}=U_{23}=0$, 
then we come back to the mean-field Lagrangian (\ref{mfL}) 
and the discussions above can be reused. 
Here, we note that in general one can take 
an arbitrary relative angle between these directed mean fields, 
i.e., taking $A_{1}\neq 0$ and $A_{2}\neq 0$ in addition to $A_3$ and $U_{12}$, 
which breaks all rotational symmetries, 
but the stationary condition 
for the thermodynamic potential should prefer phases with 
the maximal residual symmetry, i.e., $A_{1}=A_{2}=0$.  

\begin{figure}[htbp]
	\begin{center}
		\includegraphics[height=7.0cm]{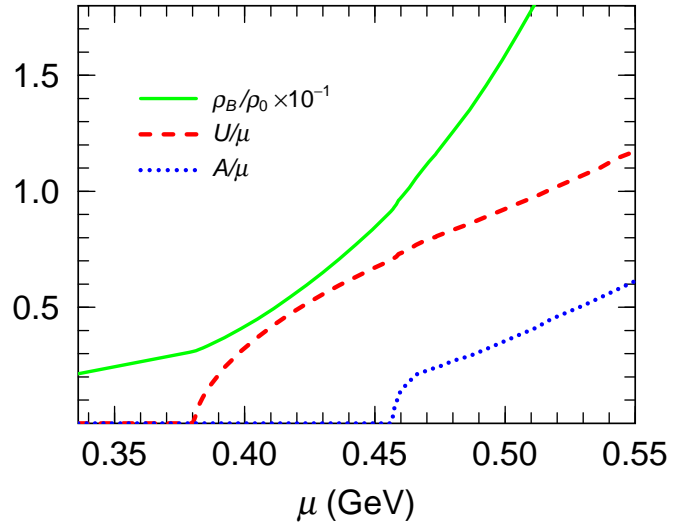}
		%    \begin{tabular}{cc}
		% \resizebox{50mm}{!}{\includegraphics{nemo1.eps}} & 
		% \resizebox{50mm}{!}{\includegraphics{nemo2.eps}} \\ 
		% \resizebox{50mm}{!}{\includegraphics{nemo2.eps}} & 
		% \resizebox{50mm}{!}{\includegraphics{nemo1.eps}} \\ 
		%    \end{tabular}
\caption{The baryon-number density $\rho_B/\rho_0$, $A/\mu$, and $U/\mu$ as functions of quark chemical potential $\mu$ 
for $g_A=73.70$ (GeV$^{-2}$) and $g_U=45.35$ (GeV$^{-2}$). 
$\rho_0=0.15$ fm$^{-3}$ is the normal nuclear density. 
  }
		\label{fig4}
	\end{center}
\end{figure}

Now we demonstrate the change of phase structure with the chemical potential by numerical calculations. 
In Fig.~\ref{fig4} we plot 
the mean fields and the baryon-number density 
as functions of the quark chemical potential $\mu$, 
where numerical values of the coupling constants are fixed 
so as to reproduce the magnetic field expected at the core of magnetars \cite{Bandyopadhyay:1997bcp,Casali:2014ccm}:
$g_A=73.70$ GeV$^{-2}$ and $g_U=45.35$ GeV$^{-2}$, 
which are of the same order of coupling constants 
used in \cite{Maedan:2006ib, Maruyama:2017mqv}, 
and are an order of magnitude larger  than  
in \cite{TPPY12, Matsuoka:2016spx,Morimoto:2018pzk}. 
We will give details of the magnetic field estimation around (\ref{mag1}) in the last section. 
Since all quantities can be scaled by the quark chemical potential, 
only which brings the energy scale in the system, 
the corresponding trajectory in the phase diagram (Fig.~\ref{fig3}) 
should be a straight line, i.e., $g_A/g_U=73.7/45.35=13/8$.  
The numerical result implies that the phase boundaries correspond to continuous phase transitions, 
and as discussed above the AV mean field starts to get finite inevitably at a point where the T mean field is already finite. 
%%%%%%%%%%%%%%%%%%%%%%%%%%%%%%%%%%%%%%%%%%%%%%%%%%%%%%%%%%%%%
%%%%%%%%%%%%%%%%%%%%%%%%%%%%%%%%%%%%%%%%%%%%%%%%%%%%%%%%%%%%%%

\subsection{Low energy modes}
Once the phase structure is determined, 
the effective degrees of freedom are low energy excitations. 
Here, we discuss the Nambu-Goldstone modes upon the spin polarized phases. 
The symmetry $G$ of the system under the strong interaction 
and at a finite quark chemical potential is given by 
\begin{equation}
G=SO(3)_{\rm rot} \otimes SU(2)_{A} \otimes SU(2)_{V} \otimes U(1)_B \otimes SU(3)_c, 
\end{equation}
where  $SO(3)_{\rm rot}$ represents the spatial rotation,
$SU(2)_{A,V}$ the axial and vectorial decomposition of the chiral flavor $SU(2)_{L,R}$ rotations, 
$U(1)_B$ the baryon-number symmetry, and $SU(3)_c$ the color gauge  symmetry. 
Broken symmetries can be read off from the transformation of the order parameters: 
For $SU(2)_A$ chiral rotations, 
i.e., $\psi \rightarrow e^{i\frac{1}{2}\gamma_5 \tau_a \pi^a}\psi$ 
where $\tau_{a=1,2,3}$ the Pauli matrices in the isospin space, 
\beq
\ldk \gamma_5 \tau_a, \gamma_0 \sigma_{12}\rdk \neq 0, \quad
\ldk \gamma_5 \tau_a, \gamma_0 \gamma_5 \gamma_3 \rdk = 0 
\eeq 
and for the spatial rotations of the spinor, 
i.e., $\psi \rightarrow e^{i\Sigma_k \theta_k}\psi$ 
where $\Sigma_k=\frac{1}{2} \sigma_{ij}\epsilon^{ijk}$ 
with $\epsilon^{123}=1$ being the anti-symmetric T, 
\beq
&&\ldk \Sigma_3, \gamma_0 \sigma_{12}\rdk = 0, \  
\ldk \Sigma_{1,2}, \gamma_0 \sigma_{12}\rdk =2i \gamma_0 \sigma_{13,23},
\\  
&&
\ldk \Sigma_3, \gamma_0 \gamma_5 \gamma_3 \rdk = 0,  \ 
\ldk \Sigma_{1,2}, \gamma_0 \gamma_5 \gamma_3 \rdk = \mp 2i\gamma_0 \gamma_5 \gamma_{2,1}. 
\eeq 
The commutations for the other generators of $G$ are vanishing. 
Thus, in the spontaneous spin polarized phases the residual symmetry $H$ becomes 
\beq
H=SO(2)_{\rm rot} \otimes SU(2)_{V}  \otimes U(1)_B \otimes SU(3)_c, 
\eeq
where $SO(2)_{\rm rot}$ reflects the invariance under a rotation around the $z$ axis. 
Note that the above symmetry breaking pattern is the same for the two spin polarized phases, 
i.e., for  ($A=0, U\neq0$) and  ($A\neq 0, U\neq0$) phases. 

Now we examine the expectation values of commutators 
among generators of the broken symmetries, 
i.e.,  
the generators of spatial rotations, 
and those of the $SU(2)_A$, defined respectively by 
\beq
\Sigma_{k}(x) &\equiv& %\int_{\bf x}
\psi^\dagger (x)\frac{1}{2}\sigma_{ij}\epsilon^{ijk} \psi(x)
=-%\int_{\bf x} 
\bar{\psi}(x) \gamma_5 \gamma_k \psi(x), \\ 
%%%%%%%
Q_{b}(x) &\equiv& %\int_{\bf x}
\psi^\dagger (x) \frac{1}{2}i\gamma_5\tau_b \psi(x), 
\eeq
where  $k=1,2$ represent spatial directions of $x, y$, 
and  $b=1,2,3$ the $SU(2)$ flavor triplet. 
We obtain the expectation values of commutators, 
\beq
\left\langle \ldk \Sigma_1(x), \Sigma_2(y) \rdk\right\rangle
&=& i\left\langle \Sigma_3(x) \right\rangle \, V \delta^{(3)}(x-y)
\nn
&=& -iA/g_A \, V \delta^{(3)}(x-y), 
\label{com1} 
\\
\left\langle\ldk \Sigma_i(x), Q_a(y) \rdk\right\rangle
&=& 
%\label{com2}  \\&&\quad 
\left\langle\ldk Q_a(x), Q_b(y) \rdk\right\rangle=0, 
\label{com3} 
\eeq
where $V$ is the volume of the system. 
From the above results 
we can classify the number of NG bosons and their dispersion relations 
\cite{Watanabe:2012hr, Hidaka:2012ym}: 
In the ($A=0, U\neq0$) phase, 
there appear two NG bosons (type I) associated with broken spatial rotational symmetries,  
and their dispersion relations is linear in momentum for low energies as
\beq
p_0=c |\vcp|, 
\eeq
where $c$ is a coefficient (sound velocity), 
while  in the ($A\neq 0, U\neq0$) phase, 
there is only single NG boson (type II) associated with broken spatial rotational symmetries 
since Eq.~(\ref{com1}) implies the broken generators $\Sigma_1$ and $\Sigma_2$ are canonically conjugate, 
i.e., there are not independent, and its dispersion relation be quadratic as
\beq
p_0=\tilde{c} \vcp^2 
\eeq
with a different coefficient $\tilde{c}$. 
In addition, 
there must be three NG bosons (type I) associated with the broken $SU(2)_A$ flavor symmetries, which have a linear dispersion.  
The $U(1)_A$ symmetry is broken in reality by quantum effect, 
thus no associated NG boson exists.  
These NG bosons are relevant degrees of freedom for low energy dynamics, e.g.,  
in scattering processes with photons and neutrinos 
\cite{Bahcall:1965zzb,Friman:1978zq,Reddy:1997yr,Tatsumi:2014cea}.

%%%%%%%%%%%%%%%%%%%%%%%%%%
\section{Summary and outlook}
We have studied the interplay between the axial-vector (AV) and tensor (T) mean fields 
for a possible spontaneous spin polarization of the strongly interacting matter 
at finite baryon-number chemical potentials and at zero temperature in the chiral limit. 
It is found that as the chemical potential increases 
the T mean field $U$ becomes finite first at a critical point, 
and breaks the chiral symmetries as well as the spatial rotation symmetries, 
while the AV mean field $A$ becomes finite only in the region of a finite $U$
because the finiteness of $A$ requires the chiral symmetry to be broken. 
All these phase boundaries correspond to continuous phase transitions. 
Furthermore, we classified the Nambu-Goldstone modes associated with broken rotational symmetries: 
there appear type I (II) NG modes with a linear (quadratic) dispersion in the phase of  $U\neq0$ and $A=0$ ($U\neq0$ and $A\neq 0$).  

In relation to the magnetic field generated by these mean fields, 
we can estimate its strength as a magnetic moment density 
$M_{mag}$ for the isospin saturated $u$-$d$ quark matter, 
\beq
M_{mag}=\lk \frac{2}{3}\bar{n}_u-\frac{1}{3}\bar{n}_d\rk \frac{e\hbar}{2m_q}
\frac{\langle \bar{\psi} \sigma_{12} \psi \rangle}{\langle \bar{\psi} \gamma_0 \psi \rangle} 3\rho_B, 
\label{mag1}
\eeq
which amounts to $M_{mag}\, \mu_0 \simeq 1.1\times 10^{18}$ Gauss 
for the quark mass $m_q=5$ MeV/C$^2$, $\mu_0$ the vacuum permeability, 
and $\bar{n}_u=\bar{n}_d$ the fraction of u,d-quark numbers $\bar{n}_u+\bar{n}_d=1$. 
The spin average 
$\frac{\langle \bar{\psi} \sigma_{12} \psi \rangle}{\langle \bar{\psi} \gamma_0 \psi \rangle}=0.22$  
is extracted from the extended NJL model (\ref{enjl}) calculated 
at $\mu=0.42$ GeV in Fig.~\ref{fig4}, 
which gives the baryon-number density $\rho_B =5.54 \, \rho_0$  
with $\rho_0=0.15\, {\rm fm}^{-3}$ being the normal nuclear density, 
high enough to expect the chirally restored quark matter \cite{Klevansky1}. 
Although the in-medium permeability may be far from the vacuum one, 
the magnetic field estimated above is almost of the same order of magnitude expected at the core of magnetars, 
where the quark matter is assumed to develop 
\cite{Bandyopadhyay:1997bcp,Casali:2014ccm}. 
Here it should be noted that 
the inside of neutron stars is isospin asymmetric due to the charge neutrality 
and the beta equilibrium conditions, 
i.e., $\bar{n}_d>\bar{n}_u$, which may lead to a big reduction of the magnetic field in the present study. 
Even in such a situation, since the strong interaction is isospin symmetric, 
we can similarly consider isovector-type spin polarizations, 
e.g., $\langle \bar{\psi} \tau_3 \sigma_{12} \psi \rangle$ as in \cite{TPPY12, Matsuoka:2016spx}, or a linear combination of isoscalar- and isovector-type spin polarizations 
as in \cite{Maruyama:2017mqv}. 
In the isovector case, the relative sign of the spin polarization between u- and d-quarks flips due to the isospin Pauli matrix $\tau_3$, 
then the charge average is replaced as $\lk \frac{2}{3}\bar{n}_u-\frac{1}{3}\bar{n}_d\rk \rightarrow \lk \frac{2}{3}\bar{n}_u+\frac{1}{3}\bar{n}_d\rk$, 
which rather leads to an enhancement.

%Outlook
In the present study 
we have ignored the contribution from the Dirac sea in the thermodynamic potential. 
The four-fermion interactions employed here in an extended NJL model 
are non-renormalizable theory, 
so that one needs to introduce a cut off  and regularization scheme, and results may be changed quantitatively depending on them. 
However, the contribution of the Dirac sea can be absorbed (renormalized) into parameters (coupling constants of terms of the potential) 
in the mean-field approximation \cite{Scavenius:2000qd,Skokov:2010sf}, 
and more importantly in the isotropic regularization scheme in momentum space such 
as the proper-time regularization the Dirac sea itself does not support the spin polarization, 
and only the finite density effects with deformed Fermi seas 
make the spin polarization possible as a many-body effect. 
Thus, the present result does not change at least qualitatively. 

One of the other directions of further investigations is to figure out 
the finite mass effects on the spontaneous spin polarization. 
As discussed above, the chiral symmetry breaking is necessary to get 
a finite AV mean field. Once the chiral symmetry breaking terms, 
such as the current mass term, are introduced,  they make $A$ easer to get finite 
even when $U=0$,  
thus the present result may change so that there appears the $(A\neq0, U=0)$ phase for some parameter regions.  
In this respect the axial anomaly, 
which breaks the $U(1)_A$ chiral symmetry explicitly, 
may affect the spin polarization as well \cite{Eser:2015pka,Huang:2017pqe}.  
%%%
We have also ignored the chiral condensation responsible 
for the dynamical chiral symmetry breaking, assuming very large baryon-number densities. 
To find more realistic phase structure, 
it is important to investigate the interplay among the spin polarization and the other orderings, 
e.g., the homogeneous/inhomogeneous chiral condensations in the moderate density region, and the color superconductivity at high density regions. 

Although the spin polarization of the dense matter 
is not defined uniquely in the relativistic framework, 
we quantify it by AV and T mean fields in this study. 
The response to external stimulations may give an another insight into 
the spin or magnetic properties of the strongly interacting matter, 
such as susceptibilities to external magnetic fields  
\cite{Ferrer:2008dy,Ferrer:2013noa,Pal:2009aj,Shovkovy:2012zn,Morimoto:2018pzk} 
and spatial rotations \cite{Vilenkin1,Ebihara:2016fwa,Chernodub:2016kxh,Liu:2017zhl}, both of which are related to neutron star physics.

%%%%%%%%%%%%%%%%%%%%%%%%%%%%%%%%%%%%%%%%%%%%%
Acknowledgement: 
T.M. and E.N. are supported by Grants-in-Aid for Scientific Research 
from the Ministry of Education, Science and Culture of Japan 
through Grant No.~16K05360 and No.~17K05445 provided by JSPS. 
%%%%%%%%%%%%%%%%%%%%%%%%%%%%%%%%%%%%%%%%%%%%%

%%%%%%%%%%%%%%%%%%%%%%%%
\appendix

\section{Integration up to Fermi surfaces}
The explicit result of the integration (\ref{omegaF})
up to th Fermi surface for $s=-1$  is given by 
\beq
&&\int_k \varepsilon_{k,-1} \theta\lk \mu -\varepsilon_{k,-1}\rk 
%%%%%%%
\nn
&=&
\frac{2}{8\pi^2} 
\int_{z_1}^{z_2}  {\rm d}k_z
\lk \frac{\varepsilon_{k,-1}^2}{6}
+\frac{y+k_z^2+A^2-U^2}{2}    \rk \varepsilon_{k,-1}
\nn
&&
-\frac{k_z^2\lk A^2-U^2\rk }{U} 
\nn
&&
\times \left. \ln\lk \sqrt{yU^2+A^2(k_z^2+U^2)}-U^2+U\varepsilon_{k,-1} \rk
\right|_{y_1}^{y_2}, 
\nn 
\eeq
where 
\beq
y_2&=&-A^2-k_z^2+U^2+\mu^2+2\sqrt{k_z^2(A^2-U^2)+U^2\mu^2}, 
%%%%%%%%%
\\
y_1&=&\ldk -A^2-k_z^2+U^2+\mu^2-2\sqrt{k_z^2(A^2-U^2)+U^2\mu^2}\rdk 
\nn
&&
\times
\theta\lk U-\sqrt{A(A+\mu)} \rk  
\nn
&&
\times \theta\lk k_z-\theta(A-U+\mu) \sqrt{(A-U+\mu)(A+U+\mu)} \rk, 
\nn 
%%%%%%%%%%
\\
&&
\nn
z_2&=& \sqrt{\lk A-U+\mu\rk\lk A+U+\mu\rk}\theta\lk \sqrt{A(A+\mu)}-U\rk 
\nn
&&+\frac{U\mu}{\sqrt{U^2-A^2}} \theta\lk U-\sqrt{A(A+\mu)}\rk, 
%%%%%%%%
\\
z_1&=& \sqrt{\lk A-U-\mu\rk\lk A+U-\mu\rk}\theta\lk A-U-\mu\rk. 
\eeq

The integration up to Fermi surface for $s=1$ is given by 
\beq
&&\int_k \varepsilon_{k,1} \theta\lk \mu -\varepsilon_{k,1}\rk 
\nn
&=&
\frac{2}{8\pi^2} 
\int_{0}^{z_3}  {\rm d}k_z
\lk \frac{\varepsilon_{k,1}^2}{6}
+\frac{y+k_z^2+A^2-U^2}{2}    \rk \varepsilon_{k,1}
\nn
&&
+\frac{k_z^2\lk A^2-U^2\rk }{U} 
\nn
&&
\times \left. \ln\lk \sqrt{yU^2+A^2(k_z^2+U^2)}+U^2+U\varepsilon_{k,1} \rk
\right|_{0}^{y_3}, 
\nn 
\eeq
where 
\beq
y_3&=&\ldk -A^2-k_z^2+U^2+\mu^2-2\sqrt{k_z^2(A^2-U^2)+U^2\mu^2}\rdk 
\nn
&&
\times \theta\lk \mu-U-A \rk,  
%%%%%%%%%%
\\
&&
\nn
z_3&=& \sqrt{\lk A-U-\mu\rk\lk A+U-\mu\rk}\, \theta\lk \mu-U-A \rk. 
\eeq

%%%%%%%%%%%%%%%%%%%%%%
\section{The potential curvature with respect to $A$}
We calculate the second derivative of the thermodynamic potential with respect to $A$,  
\beq
&&\partial_A^2 \int_k \varepsilon_{k,s} \theta\lk \mu -\varepsilon_{k,s}\rk 
\nn  
&=&
\int_k \partial_A^2\varepsilon_{k,s} \theta\lk \mu -\varepsilon_{k,s}\rk 
-\int_k \lk \partial_A\varepsilon_{k,s} \rk^2 \delta\lk \mu -\varepsilon_{k,s}\rk 
\nn 
&&- \partial_A \int_k \varepsilon_{k,s} \, \partial_A\varepsilon_{k,s} \delta\lk \mu -\varepsilon_{k,s}\rk, 
%%%%%%%%%%
\eeq
then taking the limit of  $A\rightarrow 0$ and noting that
$\partial_A \varepsilon_{k,s}|_{A\rightarrow 0}=0$,  
we obtain 
\beq
&&\left. 
\partial_A^2 \int_k \varepsilon_{k,s} \theta\lk \mu -\varepsilon_{k,s}\rk 
\right|_{A\rightarrow 0}
\nn
&=& 
\left. \int_k \partial_A^2\varepsilon_{k,s} \theta\lk \mu -\varepsilon_{k,s}\rk 
-\mu\, \partial_A \int_k  \, \partial_A\varepsilon_{k,s} \delta\lk \mu -\varepsilon_{k,s}\rk
\right|_{A\rightarrow 0}, 
\nn 
\label{2ndcal1}
\eeq
the second term of which cancels out with the second derivative of the density contribution, 
\beq
-\mu\, \partial_A^2 \int_k\theta\lk \mu -\varepsilon_{k,s}\rk 
&=& 
\mu\, \partial_A \int_k \partial_A\varepsilon_{k,s} \delta\lk \mu -\varepsilon_{k,s}\rk.  
\nn 
\eeq
%%%%%%%%%%%%
Then the first term in (\ref{2ndcal1}) can be calculated analytically,  for $s=-1$, to be 
\beq
&&
\left. \int_k
\partial_A^2 \varepsilon_{-1} \theta(\mu- \varepsilon_{-1} ) 
\right|_{A\rightarrow0}
%%%%%%%%
\nn 
&=&
\int \frac{{\rm d}z}{(2\pi)^2U}  \int {\rm d}\rho 
\frac{U\lk \rho-U\rk-z^2}{\sqrt{(U-\rho)^2+z^2}}
\theta\lk \mu- \sqrt{(U-\rho)^2+z^2}\rk
%%%%%%%%
\nn 
&=&
\frac{\mu^3}{12\pi^2 U}
\ldk 
\frac{U}{\mu^2} \sqrt{\mu ^2-U^2} 	  
+\cot^{-1}\left(\frac{U}{\sqrt{\mu^2-U^2}}\rk \right.
\nn 
&&
+\left.\frac{2U^3}{\mu^3} \log\lk\frac{U}{\mu +\sqrt{\mu^2-U^2}}\rk
\rdk \theta\lk \mu-U \rk-\frac{\mu ^3}{12\pi U}, 
\eeq
while for $s=1$, 
\beq
&&\int_k \left. 
\partial_A^2 \varepsilon_s \theta(\mu- \varepsilon_s) 
\right|_{A\rightarrow0}
%%%%%%%%
\nn 
&=&
\frac{1}{(2\pi)^2U} \int  {\rm d}z \int {\rm d}\rho 
\frac{U\lk \rho+U\rk+z^2}{\sqrt{(U+\rho)^2+z^2}}
\nn
&&
\times 
\theta\lk \mu- \sqrt{(U+\rho)^2+z^2}\rk
%%%%%%%%
\nn 
&=&
	\frac{\mu^3}{12\pi^2 U} 
	\ldk 
     \frac{U}{\mu^2} \sqrt{\mu^2-U^2}
     +\cot^{-1}\lk\frac{U}{\sqrt{\mu^2-U^2}}\rk\right. 
 \nn
 &&   
     +\left. \frac{2U^3}{\mu^3} \log\lk\frac{U}{\mu +\sqrt{\mu^2-U^2}}\rk
	\rdk \theta\lk \mu- U\rk. 
\nn 
\eeq
Finally the 2nd derivative of $\Omega$ at $A=0$ reduces to  
\beq
&&\left. \partial_A^2\Omega\right|_{A\rightarrow 0}
\nn 
&=& 
N_d\sum_s \int_k \left. 
\partial_A^2 \varepsilon_s \theta(\mu- \varepsilon_s) 
\right|_{A\rightarrow0} +\frac{2}{g_A}
%%%%%%%%%%%%%%%%
\nn 
&=&
\frac{N_d\mu^3}{6\pi^2U}
\ldk 
\frac{U}{\mu^2} \sqrt{\mu^2-U^2}
+\cot^{-1}\lk\frac{U}{\sqrt{\mu^2-U^2}}\rk \right.
\nn
&&
+\left. \frac{2U^3}{\mu^3} \log\lk\frac{U}{\mu +\sqrt{\mu^2-U^2}}\rk
\rdk \theta\lk \mu- U\rk 
\nn
&&
-\frac{N_d}{12\pi}\frac{\mu^3}{U}+\frac{2}{g_A}. 
\label{2nddrivF}
\eeq

%%%%%%%%%%%%%%%%%%%%
\section{Thermodynamic potential of $U$ at $A=0$ and determination of phase boundary}
%%%%%%%%%%%%%%%%%%%%
The thermodynamic potential at $A=0$ is analytically obtained as 
\beq
&&\lk \frac{N_d}{24\pi^2}\rk^{-1} \Omega[0,U]
%%%
\nn 
&=&
\ldk -\mu  \sqrt{\mu ^2-U^2} \left(2 \mu ^2+3 U^2\right)\right.
\nn
&&
-4 \mu ^3 U \tan ^{-1}\left(\frac{U}{\sqrt{\mu^2-U^2}}\right) 
\nn
&&
+\left. U^4 \log \left(\frac{\mu +\sqrt{\mu ^2-U^2}}{U}\right)
\rdk \theta(\mu-U)
%%%
\nn 
&&
-2\pi \mu^3U\theta(U-\mu)
+\frac{U^2}{g_U}\lk \frac{N_d}{24\pi^2}\rk^{-1}. 
\label{potU}
\eeq 
%%%%%%%%%%%%%%%%%%%%%%%%%%%%%%%%%%%%%%%%%%%%%%%%%%%%%%%%%%%%%
\begin{figure}[htbp]
	\begin{center}
		\includegraphics[height=5.3cm]{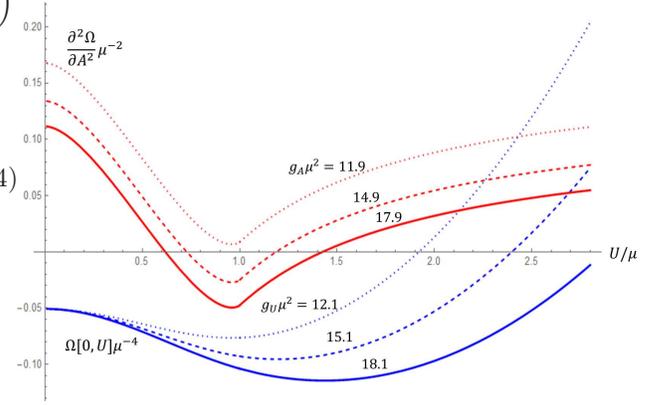}
		\caption{
The 2nd derivative (\ref{2nddrivF}) for $g_A\mu^2=g_{A1}\mu^2+2.5=14.9$ (red dashed) 
and the potential  (\ref{potU})  for $g_U\mu^2=g_{U1}\mu^2+2.5=15.1$ (blue dashed)  
are plotted as functions of $U$ in the unit of $\mu=1$. 
Red and blue solid (dotted) curves correspond to    
$g_A\mu^2=g_{A1}\mu^2+5.5=17.9$ and $g_U\mu^2=g_{U1}\mu^2+5.5=18.1$ ($g_A\mu^2=g_{A1}\mu^2-0.5=11.9$ and $g_U\mu^2=g_{U1}\mu^2-0.5=12.1$), 
respectively. }
		\label{fig6}
	\end{center}
\end{figure}
%%%%%%%%%%%%%%%%%%%%%%%%%%%%%%%%%%%%%%%%%%%%%%%%%%%%%%%%%%%%%%
The potential shape is shown in Fig.~\ref{fig6} together with the second derivative (\ref{2nddrivF})
for some values of $g_A$ and $g_U$. 
As shown in the figure,  
the value of tensor mean field at the potential minimum, $U_{\rm min}$, 
increases monotonically with $g_U$, and 
zero's of the second derivative (\ref{2nddrivF})  gives two intersection points with the abscissa for larger values of $g_A$. 
If the $U_{\rm min}$ lies in between these intersection points, the phase with $U\neq0$ and $A\neq0$ be realized. 
Therefore, the phase boundary is determined by 
searching out $g_U$, for which $U_{\rm min}$ coincides with the intersection points for a given value of $g_A (\ge g_{A1}) $. 
For smaller values of $g_A (< g_{A1})$, the intersection point disappears, 
meaning that the second derivative becomes positive everywhere. 
For larger values of $g_U$, corresponding to $U_{min}\ge \mu$, 
the phase boundary is determined simply by $g_A=g_U$. 

%%%%%%%%%%%%%%%%%%%%%%%%%%%%%%%%%%%%%%%%%%%%%%%%%%%%%%%%%%%
%\bibliography{}
%%%%%%%%%%%%%%%%%%%%%%%%%%%%%%%%%%%%%%%%%%%%%%%%%%%%%%%%%%%

%%%%%%%%%%%%%%%%%%%%%%%%

\end{document}